\documentclass[pre,tighten]{revtex4}

\usepackage{amssymb,amsmath}

\usepackage{epsfig}

\usepackage{graphicx}
\usepackage{color}

\begin{document}
\title{Rheological properties for inelastic Maxwell mixtures under shear flow}
\author{Vicente Garz\'{o}\footnote[1]{Electronic address: vicenteg@unex.es;
URL: http://www.unex.es/eweb/fisteor/vicente/}}
\affiliation{Departamento de F\'{\i}sica, Universidad de Extremadura, E-06071 Badajoz, Spain}
\author{Emmanuel Trizac\footnote[2]{Electronic address: trizac@lptms.u-psud.fr;
URL: http://www.lptms.u-psud.fr/membres/trizac/}} \affiliation{Laboratoire de Physique
Thé\'eorique et Mod\`eles Statistiques (CNRS UMR 8626), B$\hat{a}$timent 100,
Universit\'e Paris-Sud, 91405 Orsay cedex, France}

\begin{abstract}
The Boltzmann equation for inelastic Maxwell models is considered to determine
the rheological properties in a granular binary mixture in the simple shear flow state.
The transport coefficients (shear viscosity and viscometric functions) are {\em
exactly} evaluated in terms of the coefficients of restitution, the (reduced) shear
rate and the parameters of the mixture (particle masses, diameters and concentration).
The results show that in general, for a given value of the coefficients of restitution,
the above transport properties decrease with increasing shear rate.
\end{abstract}

\date{\today}
\maketitle

\section{Introduction}
\label{sec1}

It is well-recognized that granular matter can be modeled by a fluid of hard spheres with inelastic
collisions. In the simplest version, the grains are assumed to be smooth so that the inelasticity is only
accounted for by a constant coefficient of normal restitution. For sufficiently low-densities, the
(inelastic) Boltzmann equation has been solved by means of the Chapman-Enskog method \cite{CC70} and
the Navier-Stokes transport coefficients have been obtained in terms of the coefficient of restitution
\cite{BDKS98}. Moreover, some analytical results in far from equilibrium situations have been also reported
in the literature for inelastic hard spheres  \cite{IHS}. However, due to the complex mathematical structure of the Boltzmann collision operator, it is generally not possible to get exact results from the Boltzmann equation for inelastic hard spheres and consequently, most of the analytical results have been derived by using approximations and/or kinetic models.

As in the elastic case, a possible way to partially overcome the above difficulties is to consider the
so-called inelastic Maxwell models (IMM), where the collision rate is independent of the relative velocity
of the two colliding particles \cite{Maxwell}. Thanks to this property, nonlinear transport properties can
be {\em exactly} obtained in some particular problems \cite{G03,TK03,SG07} for IMM without introducing additional
and sometimes uncontrolled approximations.  In addition, apart from their academic interest, it has also
been shown that in some cases the results derived from IMM compare well with those obtained for inelastic hard spheres \cite{G03} and even recent experiments \cite{exp05} for magnetic grains with dipolar interactions are well described by IMM. All these results stimulate the use of this interaction model as a toy model to characterize the influence of the inelasticity of collisions on the physical properties of granular fluids.

The aim of this paper is to determine the rheological properties
(shear stress and normal stress differences) in a binary mixture described by the Boltzmann equation  for IMM and subjected
to the simple or uniform shear flow (USF). This state is perhaps one of the most widely studied states in
granular gases \cite{Go03}. At a macroscopic level, the USF is characterized by constant partial densities
$n_r$, a uniform granular temperature $T$, and a linear velocity profile $u_x=ay$, where $a$ is the constant
shear rate. Under these conditions, the mass and heat fluxes vanish by symmetry and the pressure 
tensor $P_{ij}$ is the only relevant flux in the problem. Conservation of momentum implies
$P_{i,y}=\text{const}$ while the energy balance equation reads
\begin{equation}
\label{1.1}
\frac{\partial}{\partial t} T=-\frac{2a}{d n} P_{xy}-T \zeta,
\end{equation}
where $d$ is the dimensionality of the system ($d=3$ for spheres and $d=2$ for disks) and $\zeta$ is the
inelastic cooling rate. Equation (\ref{1.1}) clearly shows that the temperature changes over time due to
two competing effects: the viscous heating term $a P_{xy}$ and the inelastic collisional cooling term
$\zeta T$. While the first term is inherently positive (since $P_{xy}<0$), the second term is negative
since $\zeta>0$.
Depending on the initial condition, one of the effects prevails over the other one so that the temperature
either decreases or increases in time, until a steady state is reached for sufficiently long times.
Given that in this steady state the (reduced) shear rate is enslaved to the coefficients of restitution,
the problem is inherently non-Newtonian (and so, beyond the scope of the
Navier-Stokes description) in highly inelastic granular gases \cite{SGD04}.

As in a previous paper for a single gas \cite{SG07}, the rheological properties of the granular binary mixture are given in terms of a collision frequency $\nu_0$, which in principle can be freely chosen. Here we will consider two classes of IMM: (i) a collision frequency $\nu_0$ independent of temperature (Model A) and (ii) a collision frequency $\nu_0(T)$ monotonically increasing with temperature (Model B). Model A is closer to the original model of Maxwell molecules for elastic gases \cite{TM80,GS03} while Model B with $\nu_0(T)\propto \sqrt{T}$ is closer to inelastic hard spheres. The possibility of having a general temperature dependence of $\nu_0(T)$ for inelastic repulsive models has been recently introduced by Ernst and co-workers \cite{ETB06}. As will be shown later, Models A and B lead to the same results in the steady state limit. In particular, the reduced shear rate $a^*=a/\nu_0$ in the steady state is a {\em universal} function $a_s^*(\alpha_{ij})$ of the coefficients of restitution $\alpha_{ij}$ and the parameters of the mixture. However, since $a^*$ does not change in time for Model A, a steady state does not exist except for the specific value $a^*=a_s^*(\alpha_{ij})$. Consequently, a non-Newtonian hydrodynamic regime (where $a^*$ and $\alpha_{ij}$ are independent parameters) is reached in the long-time limit where the combined effect of both control parameters on the rheological properties can be studied analytically for Model A. This is an interesting new added value of this simple model.

The plan of the paper is as follows. We first introduce in Section \ref{sec2} the Boltzmann equation framework
for IMM; collisional moments are worked out.
In Section \ref{sec3} we will introduce driving through a macroscopic shear, and consider
in particular the USF problem. We shall subsequently focus on rheological properties in Section
\ref{sec4}, where our results for the non-linear shear viscosity and viscometric functions will be reported.
Finally, conclusions will be drawn in Section \ref{sec5}.

\section{The Boltzmann equation for IMM. Collisional moments}
\label{sec2}

Let us consider a binary mixture of inelastic Maxwell gases at low density. In the
absence of external forces, the set of nonlinear  Boltzmann equations for the mixture
reads
\begin{equation}
\label{2.1}
\left(\frac{\partial}{\partial t}+{\bf v}\cdot \nabla \right)f_{r}
({\bf r},{\bf v};t)
=\sum_{s}J_{rs}\left[{\bf v}|f_{r}(t),f_{s}(t)\right] \;,
\end{equation}
where $f_r({\bf r},{\bf v};t)$ is the one-particle distribution function of species $r$ ($r=1,2$) and
the Boltzmann collision operator $J_{rs}\left[{\bf v}|f_{r},f_{s}\right]$ describing the scattering of
pairs of particles is
\begin{widetext}
\begin{equation}
J_{rs}\left[{\bf v}_{1}|f_{r},f_{s}\right]  =\frac{\omega_{rs}}{n_s\Omega_d}
\int d{\bf v}_{2}\int d\widehat{\boldsymbol {\sigma }}\left[ \alpha_{rs}^{-1}f_{r}({\bf r},{\bf v}_{1}',t)f_{s}(
{\bf r},{\bf v}_{2}',t)-f_{r}({\bf r},{\bf v}_{1},t)f_{s}({\bf r},{\bf v}_{2},t)\right]
\;.
\label{2.2}
\end{equation}
\end{widetext}
Here,
\begin{equation}
\label{2.4.1} n_r=\int d{\bf v} f_r({\bf v})
\end{equation}
is the number density of species $r$, $\omega_{rs}$ is an effective collision frequency
(to be chosen later) for collisions  of type $r-s$,  $\Omega_d=2\pi^{d/2}/\Gamma(d/2)$
is the total solid angle in $d$ dimensions, and $\alpha_{rs}\leq 1$ refers to the
constant coefficient of restitution  for collisions between particles of species $r$
with $s$.   In addition, the primes on the velocities denote the initial values $\{{\bf
v}_{1}^{\prime}, {\bf v}_{2}^{\prime}\}$ that lead to $\{{\bf v}_{1},{\bf v}_{2}\}$
following a binary collision:
\begin{equation}
\label{2.3}
{\bf v}_{1}^{\prime }={\bf v}_{1}-\mu_{sr}\left( 1+\alpha_{rs}
^{-1}\right)(\widehat{\boldsymbol {\sigma}}\cdot {\bf g}_{12})\widehat{\boldsymbol
{\sigma}},
\end{equation}
\begin{equation}
\label{2.3.1}
{\bf v}_{2}^{\prime}={\bf v}_{2}+\mu_{rs}\left(
1+\alpha_{rs}^{-1}\right) (\widehat{\boldsymbol {\sigma}}\cdot {\bf
g}_{12})\widehat{\boldsymbol{\sigma}}\;,
\end{equation}
where ${\bf g}_{12}={\bf v}_1-{\bf v}_2$ is the relative velocity of the colliding pair,
$\widehat{\boldsymbol {\sigma}}$ is a unit vector directed along the centers of the two colliding
spheres, and $\mu_{rs}=m_r/(m_r+m_s)$.

The effective collision frequencies $\omega_{rs}$ are independent of velocity but
depend on space an time through its dependence on density and temperature. Here, as in
previous works \cite{SG07} for monocomponent gases, we will assume that
$\omega_{rs}\propto n_s T^q$, with $q\geq 0$. The case $q=0$ (a collision frequency
independent of temperature) will be referred to as Model A while the case $q > 0$
(collision frequency monotonically increasing with temperature) will be called Model B.
Model A is closer to the original model of Maxwell molecules for elastic gases
\cite{TM80,GS03} while Model B, with $q=\frac{1}{2}$, is closer to hard spheres.

Apart from $n_r$, at a hydrodynamic level,  the relevant quantities in a binary mixture
are the flow velocity  ${\bf u}$, and the ``granular'' temperature $T$. They are
defined in terms of moments of the distribution $f_r$ as
\begin{equation}
\label{2.4}
\rho{\bf u}=\sum_r\rho_r{\bf u}_r=\sum_r\int d{\bf v}m_r{\bf v}f_r({\bf
v}),
\end{equation}
\begin{equation}
\label{2.5}
nT=\sum_rn_rT_r=\sum_r\int d{\bf v}\frac{m_r}{d}V^2f_r({\bf v}),
\end{equation}
where $\rho_r=m_rn_r$, $n=n_1+n_2$ is the total number density, $\rho=\rho_1+\rho_2$ is
the total mass density, and ${\bf V}={\bf v}-{\bf u}$ is the peculiar velocity.
Equations (\ref{2.4}) and (\ref{2.5}) also define the flow velocity ${\bf u}_r$ and the
partial temperature $T_r$ of species $r$, the latter measuring the mean kinetic energy
of species $r$. As confirmed by computer simulations \cite{computer}, experiments
\cite{exp} and kinetic theory calculations \cite{GD99}, the global granular temperature
$T$ is in general different from the partial temperatures $T_r$.

The collision operators conserve the particle number of each
species and the total momentum but the total energy is not
conserved:
\begin{equation}
\int d{\bf v}J_{rs}[{\bf v}|f_{r},f_{s}]=0\;,  \label{2.5.1}
\end{equation}
\begin{equation}
\sum_{r=1}^2\sum_{s=1}^2m_r\int d{\bf v}{\bf v}J_{rs}[{\bf
v}|f_{r},f_{s}]=0 \;, \label{2.5.2}
\end{equation}
\begin{equation}
\sum_{r=1}^2\sum_{s=1}^2m_r\int d{\bf v}V^{2}J_{rs}[{\bf v}
|f_{r},f_{s}]=-d nT\zeta \;,  \label{2.5.3}
\end{equation}
where $\zeta$ is identified as the total ``cooling rate'' due to
inelastic collisions among all species. At a kinetic level, it is also convenient to
discuss energy transfer in terms of the ``cooling rates'' $\zeta_r$ for the
partial temperatures $T_r$. They are defined as
\begin{equation}
\label{2.7} \zeta_r=\sum_s \zeta_{rs}=-\frac{1}{dn_rT_r}\sum_s\int d{\bf
v}m_rV^{2}J_{rs}[{\bf v}|f_{r},f_{s}]\;.
\end{equation}
The second equality  in (\ref{2.7}) defines $\zeta_{rs}$.
The total cooling rate $\zeta$ can be expressed in terms of the partial cooling rates $\zeta_r$ as
\begin{equation}
\label{2.8}
\zeta=T^{-1}\sum_{r=1}^2\; x_rT_r\zeta_r,
\end{equation}
where $x_r=n_r/n$ is the mole fraction of species $r$.

From Eqs.\ (\ref{2.5.1}) to (\ref{2.5.3}), the macroscopic balance equations for the mass, momentum and
energy of the binary mixture can be easily obtained. They are given by
\begin{equation}
D_{t}n_{r}+n_{r}\nabla \cdot {\bf u}+\frac{\nabla \cdot {\bf
j}_{r}}{m_{r}} =0\;,  \label{2.8.1}
\end{equation}
\begin{equation}
D_{t}{\bf u}+\rho ^{-1}\nabla \cdot {\sf P}={\bf 0}\;,
\label{2.8.2}
\end{equation}
\begin{equation}
D_{t}T-\frac{T}{n}\sum_{r=1}^2\frac{\nabla \cdot {\bf
j}_{r}}{m_{r}}+\frac{2}{dn} \left( \nabla \cdot {\bf q}+{\sf
P}:\nabla {\bf u}\right) =-\zeta \,T\;. \label{2.8.3}
\end{equation}
In the above equations, $D_{t}=\partial _{t}+{\bf u}\cdot \nabla $
is the material derivative,
\begin{equation}
{\bf j}_{r}=m_{r}\int d{\bf v}\,{\bf V}\,f_{r}({\bf v}),
\label{2.12}
\end{equation}
is the mass flux for species $r$ relative to the local flow,
\begin{equation}
{\sf P}=\sum_{r}\,\int d{\bf v}\,m_{r}{\bf V}{\bf V}\,f_{r}({\bf  v}),
\label{2.13}
\end{equation}
is the total pressure tensor, and
\begin{equation}
{\bf q}=\sum_{r}\,\int d{\bf v}\,\frac{1}{2}m_{r}V^{2}{\bf V}
\,f_{r}({\bf v})
\label{2.14}
\end{equation}
is the total heat flux.

The main advantage of the Boltzmann equation for Maxwell models (both elastic and
inelastic) is that the (collisional) moments of $J_{rs}[f_r,f_s]$ can be exactly
evaluated in terms of the moments of $f_r$ and $f_s$ without the explicit knowledge of
both distribution functions \cite{TM80}. This property has been recently exploited
\cite{GS07} to obtain the detailed expressions for all the second-, third- and
fourth-degree collisional moments for a monodisperse gas. In the case of a binary
mixture, only the first- and second-degree collisional moments have been also
explicitly obtained. In particular \cite{G03},
\begin{widetext}
\begin{eqnarray}
\label{2.16}
\int d{\bf v} m_r {\bf V} {\bf V}J_{rs}[f_r,f_s]&=&
-\frac{\omega_{rs}}{\rho_sd}\mu_{sr}(1+\alpha _{rs})\left\{2\rho_s{\sf P}_r-\left(
{\bf j}_r{\bf j}_s+{\bf j}_s{\bf j}_r\right)\right. \nonumber\\
& &-\frac{2}{d+2}\mu_{sr}(1+\alpha _{rs})\left[\rho_s{\sf P}_r+\rho_r{\sf P}_s-
\left({\bf j}_r{\bf j}_s+{\bf j}_s{\bf j}_r\right)\right.\nonumber\\
& & \left.\left.
 +\left[\frac{d}{2}\left(\rho_rp_s+\rho_sp_r\right)-{\bf j}_r\cdot {\bf j}_s\right]\openone
\right]\right\},
\end{eqnarray}
\end{widetext}
where
\begin{equation}
\label{3.4}
{\sf P}_r=\int d{\bf v}\,m_{r}{\bf V}{\bf V}\,f_{r},
\end{equation}
$p_r=n_rT_r=\text{tr}{\sf P}_{r}/d$ is the partial pressure of species $r$, and
$\openone$ is the $d\times d$ unit tensor. It must be remarked that, in general beyond
the linear hydrodynamic regime (Navier-Stokes order), the above property of the
Boltzmann collision operator is not sufficient to exactly solve the hierarchy of moment
equations due to the free-streaming term of the Boltzmann equation. Nevertheless, there
exist some particular situations (such as the simple shear flow problem) for which the
above hierarchy can be recursively solved.

The cooling rates $\zeta_{rs}$ defined by Eq.\ (\ref{2.7}) can be easily obtained from Eq.\ (\ref{2.16}) as
\begin{equation}
\label{2.15}
\zeta_{rs}=\frac{2\omega_{rs}}{d}\mu_{sr}(1+\alpha _{rs})\left[1-\frac{\mu_{sr}}{2}(1+\alpha_{rs})
\frac{\theta_r+\theta_s}{\theta_s}
+\frac{\mu_{sr}(1+\alpha _{rs})-1}{d\rho_sp_r}
{\bf j}_r\cdot {\bf j}_s\right],
\end{equation}
where
\begin{equation}
\label{2.16.1} \theta_r=\frac{m_r}{\gamma_r}\sum_{s}m_s^{-1},
\end{equation}
and $\gamma_r\equiv T_r/T$ . Equation (\ref{2.15}) provides the relationship between
the collision frequencies $\omega_{rs}$ and the cooling rates $\zeta_{rs}$. This
relationship can be used to fix the explicit forms of $\omega_{rs}$. As in previous
works on inelastic Maxwell mixtures \cite{G03,GA05}, $\omega_{rs}$ is chosen here to
guarantee that the cooling rate for IMM be the same as that of inelastic hard spheres (evaluated at the
local equilibrium approximation) \cite{GD99}. With this choice, one gets
\begin{equation}
\label{2.32}
\omega_{rs}=x_s\left(\frac{\sigma_{rs}}{\sigma_{12}}\right)^{d-1}
\left(\frac{\theta_r+\theta_s}{\theta_r\theta_s}\right)^{1/2}\nu_0, \quad \nu_0=A(q) n T^q,
\end{equation}
where the value of the quantity $A(q)$ is irrelevant for our purposes. Upon deriving
(\ref{2.32}) use has been made of the fact that the mass flux ${\bf j}_r$ vanishes in
the local equilibrium approximation. In the remainder of this paper, we will take the
choice (\ref{2.32}) for $\omega_{rs}$. The results for IMM \cite{G03} obtained with the
latter choice in the {\em steady} shear flow problem compare very well with those
theoretically obtained for inelastic hard spheres in the first Sonine approximation and by means of Monte
Carlo simulations \cite{MG02}.

\section{Uniform shear flow}
\label{sec3}

Let us assume that the binary mixture is under USF. The USF state is
macroscopically defined by constant densities $n_r$, a spatially uniform temperature
$T(t)$ and a linear velocity profile ${\bf u}(y)={\bf u}_1(y)={\bf u}_2(y)=ay
\widehat{{\bold x}}$, where $a$ is the {\em constant} shear rate. Since $n_r$ and $T$
are uniform, then ${\bf j}_r={\bf q}={\bf 0}$, and the transport of momentum (measured
by the pressure tensor) is the relevant phenomenon. At a microscopic level, the USF is
characterized by a velocity distribution function that becomes {\em uniform} in the
local Lagrangian frame, i.e., $f_r({\bf r},{\bf v};t)=f_r({\bf V},t)$. In this frame,
the Boltzmann equation (\ref{2.1}) reads \cite{GS03}
\begin{equation}
\label{3.1} \frac{\partial}{\partial t}f_1-aV_y\frac{\partial}{\partial V_x}f_1=J_{11}[f_1,f_1]+J_{12}[f_1,f_2]
\end{equation}
and a similar equation for $f_2$. Equation (\ref{3.1}) is invariant under the
transformations $(V_x,V_y)\to (-V_x,-V_y)$, $V_i\to -V_i$, with $i\neq x, y$. This
implies that if the initial state is compatible with the latter symmetry properties,
then the solution to (\ref{3.1}) has the same properties at any time $t>0$. Note that
the properties of uniform temperature and constant densities and shear rate are
enforced in computer simulations by applying the Lees-Edwards boundary conditions
\cite{LE72}, regardless of the particular interaction model considered. In the case of
boundary conditions representing realistic plates in relative motion, the corresponding non-equilibrium state
is the so-called Couette flow, where densities, temperature and shear rate are no longer uniform \cite{Tij01,VGS08}.

As said before, the rheological properties of the mixture are obtained from the
pressure tensor ${\sf P}={\sf P}_1+{\sf P}_2$, where the partial pressure tensors ${\sf
P}_r$ ($r=1,2$) are defined by Eq.\ (\ref{3.4}). The elements of these tensors can be
obtained by multiplying the Boltzmann equation (\ref{3.1}) by $m_r{\bf V}{\bf V}$ and
integrating over ${\bf V}$. The result is
\begin{widetext}
\begin{equation}
\label{3.5}
\frac{\partial}{\partial t}P_{1,ij}+a_{ik}P_{1,kj}+a_{jk}P_{1,ki}+B_{11}P_{1,ij}+B_{12}P_{2,ij}=
\left(A_{11}p_{1}+A_{12}p_2\right)\delta_{ij},
\end{equation}
\end{widetext}
where use has been made of Eq.\ (\ref{2.16}) (with ${\bf j}_r={\bf 0}$). In Eq.\
(\ref{3.5}), $a_{ij}=a\delta_{ix}\delta_{jy}$ and we have introduced the coefficients
\begin{equation}
\label{3.6}
A_{11}=\frac{\omega_{11}}{2(d+2)}(1+\alpha_{11})^2+\frac{\omega_{12}}{d+2}\mu_{21}^2(1+\alpha_{12})^2,
\end{equation}
\begin{equation}
\label{3.7}
A_{12}=\frac{\omega_{12}}{d+2}\frac{\rho_1}{\rho_2}\mu_{21}^2(1+\alpha_{12})^2,
\end{equation}
\begin{equation}
\label{3.8}
B_{11}=\frac{\omega_{11}}{d(d+2)}(1+\alpha_{11})(d+1-\alpha_{11})
+\frac{2\omega_{12}}{d(d+2)}\mu_{21}(1+\alpha_{12})
\left[d+2-\mu_{21}(1+\alpha_{12})\right],
\end{equation}
\begin{equation}
\label{3.9}
B_{12}=-\frac{2}{d}A_{12}.
\end{equation}
A similar equation can be obtained for ${\sf P}_2$, by adequate change of indices
$1\leftrightarrow 2$. The balance equation (\ref{1.1}) for the temperature  can be easily obtained
from Eq.\ (\ref{3.5}). In reduced units, Eq.\ (\ref{1.1}) can be written as
\begin{equation}
\label{3.10}
\nu_0^{-1}\frac{\partial}{\partial t}\ln T=-\zeta^*-\frac{2a^*}{d} P_{xy}^*,
\end{equation}
where $\zeta^*=\zeta/\nu_0$, $a^*=a/\nu_0$, $P_{xy}^*=P_{xy}/p$, $p=nT$ being the
hydrostatic pressure. The expression for $\zeta^*$ can be obtained from Eqs.\
(\ref{2.7}), (\ref{2.8}) and (\ref{2.15}) when one takes ${\bf j}_r={\bf 0}$. It is given by
\begin{equation}
\label{3.10.1}
\zeta^*=\frac{2}{d}\sum_{r=1}^2\sum_{s=1}^2\; x_rx_s  \left(\frac{\sigma_{rs}}{\sigma_{12}}\right)^{d-1}
\left(\frac{\theta_r+\theta_s}{\theta_r\theta_s}\right)^{1/2} \gamma_r \mu_{sr}
(1+\alpha_{sr})
\left[1-\frac{\mu_{sr}}{2}(1+\alpha_{rs})
\frac{\theta_r+\theta_s}{\theta_s}\right].
\end{equation}

As said in the Introduction, Eq.\ (\ref{3.10}) shows that the temperature changes in time due the competition of two
opposite mechanisms: on the one hand, viscous heating (shearing work) and, on the other
hand, energy dissipation in collisions. Moreover, the {\em reduced} shear rate $a^*$ is
the non-equilibrium relevant parameter of the USF problem since it measures the distance
of the system from the homogeneous cooling state. It is apparent that, except for Model
A ($q=0$), the collision frequency $\nu_0(T)\propto T^{q}$ is an increasing function of
temperature (provided $q>0$), and so $a^*(t)\propto T(t)^{-q}$ is a function of time.
Consequently, for $q\neq 0$, after
a transient regime a steady state is achieved in the long time limit when both viscous
heating and collisional cooling cancel each other and the mixture autonomously seeks
the temperature at which the above balance occurs. In this steady state, the reduced
shear rate and the coefficients of restitution are not independent parameters since
they are related through the steady state condition
\begin{equation}
\label{3.11}
a_s^*P_{s,xy}^*=-\frac{d}{2}\zeta^*,
\end{equation}
where we have called $a_s^*$ and $P_{s,xy}^*$ the steady-state values of the (reduced) shear rate and the pressure tensor. On the other hand, when $q=0$, $\partial_t a^*=0$ and the reduced shear rate
remains in its initial value regardless of the values of the coefficients of
restitution $\alpha_{rs}$. As a consequence,
in the case of Model A,
there is no steady state (unless $a^*$
takes the specific value $a_s^*$ given by the condition (\ref{3.11})) and $a^*$ and
$\alpha_{rs}$ are \emph{independent} parameters in the USF problem.
The analytical
study of the combined effect of both control parameters on the pressure tensor
is the main goal of this paper.

\section{Rheological properties}
\label{sec4}

In order to characterize the nonlinear response of the system to the action of strong shearing, it is convenient to introduce the non-linear shear viscosity $\eta^*$ and the viscometric functions $\Psi_1^*$ and $\Psi_2^*$ as
\begin{equation}
\label{3.11.1}
\eta^*(a^*)=-\frac{\nu_0}{p}\frac{P_{xy}}{a},
\end{equation}
\begin{equation}
\label{3.11.2}
\Psi_1^*(a^*)=\frac{\nu_0^2}{p}\frac{P_{xx}-P_{yy}}{a^2}, \quad
\Psi_2^*(a^*)=\frac{\nu_0^2}{p}\frac{P_{zz}-P_{yy}}{a^2}.
\end{equation}
The viscosity function $\eta^*(a^*)$ is a measure of the breakdown of the linear relationship between
the shear stress $P_{xy}$ and the shear rate (Newton's law), while the first and second viscometric
functions $\Psi_{1,2}^*(a^*)$ represent the normal stress differences. The explicit form of the above functions depends on the interaction model considered.

\subsection{Model A}

In Model A ($q=0$), the collision frequency is independent of temperature and the reduced shear rate $a^*$ is a constant. Thus, Eq.\ (\ref{3.5}) and its counterpart for ${\sf P}_2$
constitute a linear homogeneous set of coupled differential equations. In fact, it is
easy to see that the relevant elements of the partial pressure tensors are the
$xy$-elements along with the diagonal ones. As expected, the remaining elements tend to
zero in the long-time limit. Moreover, from Eq.\ (\ref{3.5}) is also easy to prove
that, for long times, $P_{r,yy}=P_{r,zz}=\cdots=P_{r,dd}$. Thus, according to Eq.\ (\ref{3.11.2}), the second viscometric function $\Psi_2^*=0$. This is a particular property of IMM since $\Psi_2^*\neq 0$ for inelastic hard spheres
\cite{MG02}, although its magnitude is always much smaller than that of $\Psi_1^*$. As a consequence, the relevant elements of the partial
pressure tensors are $P_{r,xx}=p_r-(d-1)P_{r,yy}$, $P_{r,yy}$, and $P_{r,xy}$ with $r=1,2$.

As in the monocomponent granular case \cite{SG07}, one can check that, after a
certain kinetic regime lasting a few collision times, the scaled pressure tensors
$P_{r,ij}^*=P_{r,ij}/p$ reach well-defined stationary values (non-Newtonian hydrodynamic regime), which are non-linear
functions of the (reduced) shear rate $a^*=a/\nu_0$ and the coefficients of
restitution. In terms of these scaled variables and by using matrix notation, Eq.\
(\ref{3.5}) can be rewritten as
\begin{equation}
\label{3.12}
{\cal L}{\cal P}={\cal Q},
\end{equation}
where ${\cal P}$ is the column matrix defined by the set
\begin{equation}
\label{3.13}
{\cal P}\equiv \{P_{1,xx}^*, P_{1,yy}^*, P_{1,xy}^*,P_{2,xx}^*,P_{2,yy}^*,P_{2,xy}^*\},
\end{equation}
${\cal Q}$ is the column matrix
\begin{equation}
\label{3.13.1}
{\cal Q}=\left(
\begin{array}{c}
A_{11}^*p_1^*+A_{12}^*p_2^*\\
A_{11}^*p_1^*+A_{12}^*p_2^*\\
0\\
A_{22}^*p_2^*+A_{21}^*p_1^*\\
A_{22}^*p_2^*+A_{21}^*p_1^*\\
0
\end{array}
\right),
\end{equation}
and ${\cal L}$ is the square matrix
\begin{widetext}
\begin{equation}
\label{3.14}
{\cal L}=\left(
\begin{array}{cccccc}
B_{11}^*+\lambda&0&2a^*&B_{12}^*&0&0\\
0&B_{11}^*+\lambda&0&0&B_{12}^*&0\\
0&a^*&B_{11}^*+\lambda&0&0&B_{12}^*\\
B_{21}^*&0&0&B_{22}^*+\lambda&0&2a^*\\
0&B_{21}^*&0&0&B_{22}^*+\lambda&0\\
0&0&B_{21}^*&0&a^*&B_{22}^*+\lambda
\end{array}
\right).
\end{equation}
\end{widetext}
Here, $p_r^*=p_r/p=x_r\gamma_r$, $A_{rs}^*=A_{rs}/\nu_0$ and $B_{rs}^*=B_{rs}/\nu_0$.
Moreover, on physical grounds it has been assumed that for long times the temperature
behaves as
\begin{equation}
\label{3.15}
T(t)=T(0)e^{\lambda \nu_0 t}
\end{equation}
where $\lambda$ is also a nonlinear function of $a^*$, $\alpha_{rs}$ and the parameters
of the mixture. The (reduced) total pressure tensor $P_{ij}^*=P_{ij}/p$ of the mixture is defined as
\begin{equation}
\label{3.15.1}
P_{ij}^*= P_{1,ij}^*+P_{2,ij}^*.
\end{equation}

The solution to Eq.\ (\ref{3.12}) is
\begin{equation}
\label{3.16}
{\cal P}={\cal L}^{-1}\cdot {\cal Q}.
\end{equation}
The explicit forms for $P_{1,xx}^*$, $P_{1,yy}^*$, and $P_{1,xy}^*$ can be found in the
Appendix. The corresponding expressions for species $2$ are easily obtained by
adequately changing the indices. This solution is still formal as we do not know the
shear-rate dependence of $\lambda$ and the temperature ratios $\gamma_1$ and
$\gamma_2$. These quantities must be consistently determined from the requirements
\begin{equation}
\label{3.17}
x_1\gamma_1=\frac{P_{1,xx}^*+(d-1)P_{1,yy}^*}{d},
\end{equation}
\begin{equation}
\label{3.18}
x_2\gamma_2=\frac{P_{2,xx}^*+(d-1)P_{2,yy}^*}{d},
\end{equation}
\begin{equation}
\label{3.19}
\gamma_2=\frac{1-x_1\gamma_1}{x_2},
\end{equation}
which follows from Eq.\ (\ref{2.5}). Since the collision frequencies $\omega_{rs}$ are
nonlinear functions of the temperature ratios, then it is not possible to get a closed
equation for $\lambda$ or $\gamma_r$. Thus, one has to numerically solve the set of
nonlinear equations (\ref{3.17}) and (\ref{3.18}).

Nevertheless, there are some limiting cases for which the problem can be solved analytically. For instance, in the case of mechanically equivalent particles ($m_1=m_2, \sigma_1=\sigma_2,
\alpha_{11}=\alpha_{22}=\alpha_{12}$), one gets that $\gamma_1=\gamma_2=1$ and the
partial pressure tensors $P_{r,ij}^*$ can be written as
\begin{equation}
\label{3.20} \frac{P_{1,yy}^*}{x_1}=\frac{P_{2,yy}^*}{x_2}=\frac{1}{1+2\Lambda},
\end{equation}
\begin{equation}
\label{3.20.1}
\frac{P_{1,xx}^*}{x_1} =\frac{P_{2,xx}^*}{x_2}=\frac{1+2d\Lambda}{1+2\Lambda},
\end{equation}
\begin{equation}
\label{3.20.2}
\frac{P_{1,xy}^*}{x_1}=\frac{P_{2,xy}^*}{x_2}=-\frac{\widetilde{a}}{(1+2\Lambda)^2},
\end{equation}
where
\begin{equation}
\label{3.21} \widetilde{a}=\frac{2(d+2)}{(1+\alpha)^2}a^*,
\end{equation}
and $\Lambda$ is the real root of the equation
\begin{equation}
\label{3.22} \Lambda(1+2\Lambda)^2=\frac{\widetilde{a}^2}{d},
\end{equation}
namely,
 \begin{equation}
\label{3.23}
\Lambda(\widetilde{a})=\frac{2}{3}\sinh^2\left[\frac{1}{6}\cosh^{-1}\left(1+\frac{27}{d}\widetilde{a}^2\right)\right].
\end{equation}
The parameter $\lambda$ governing the long-time behavior of the temperature can be easily obtained from Eqs.\ (\ref{3.10}) and (\ref{3.15}) as
\begin{eqnarray}
\label{3.24} \lambda&=&-\zeta^*-\frac{2 a^*}{d} P_{xy}^*\nonumber\\
&=&\frac{(1+\alpha)^2}{d+2}\Lambda-\frac{1-\alpha^2}{2d}
\end{eqnarray}
where use has been made of the result $\zeta^*=(1-\alpha^2)/2d$. Equations (\ref{3.20})--(\ref{3.24}) are the
same as those obtained for a monocomponent gas \cite{SG07,rque}. Moreover, in the absence of shear field ($a^*=0$), $P_{r,xx}^*=P_{r,yy}^*=x_r\gamma_r$ and the shear viscosity function $\eta^*=\eta_1^*+\eta_2^*$ where
\begin{equation}
\label{3.25}
\eta_1^*=\frac{x_1\gamma_1(B_{22}^*-\zeta^*)+x_2\gamma_2B_{12}^*}{(B_{11}^*-\zeta^*)(B_{22}^*-\zeta^*)-
B_{12}^*B_{21}^*},\quad
1\leftrightarrow 2.
\end{equation}
The temperature ratio $\gamma=\gamma_1/\gamma_2$ is determined from the condition
\begin{equation}
\label{3.26}
\frac{x_1}{x_2}\gamma=\frac{P_{1,xx}^*+(d-1)P_{1,yy}^*}{P_{2,xx}^*+(d-1)P_{2,yy}^*}.
\end{equation}
These results are consistent with those obtained for IMM in the Navier-Stokes regime
\cite{GA05}.

It must be remarked that, although the scaled pressure tensors $P_{r,ij}^*$ reach stationary values, the gas is not in general in a steady state since the temperature changes in time. Actually, according to Eqs.\ (\ref{3.10}) and (\ref{3.11.1}), one gets
\begin{equation}
\label{3.27}
\nu_0^{-1}\partial_t \ln T=-\zeta^*+\frac{2a^{*2}}{d}\eta^*.
\end{equation}
Equation (\ref{3.27}) shows that $T(t)$ either grows or decays exponentially.
The first situation occurs if $2 a^{*2}\eta^*>d \zeta^*$. In that case, the imposed shear rate is
sufficiently large (or the inelasticity is sufficiently low) as to make the viscous heating effect
dominate over the inelastic cooling. The opposite happens if $d \zeta^*>2 a^{*2}\eta^*$. A perfect
balance between both effects takes place when $2 a^{*2}\eta^*=d \zeta^*$.

\null\vskip 5mm
\begin{figure}[!h]
\begin{minipage}[c]{\textwidth}
\null\vskip 5mm
\includegraphics[width=0.5\textwidth]{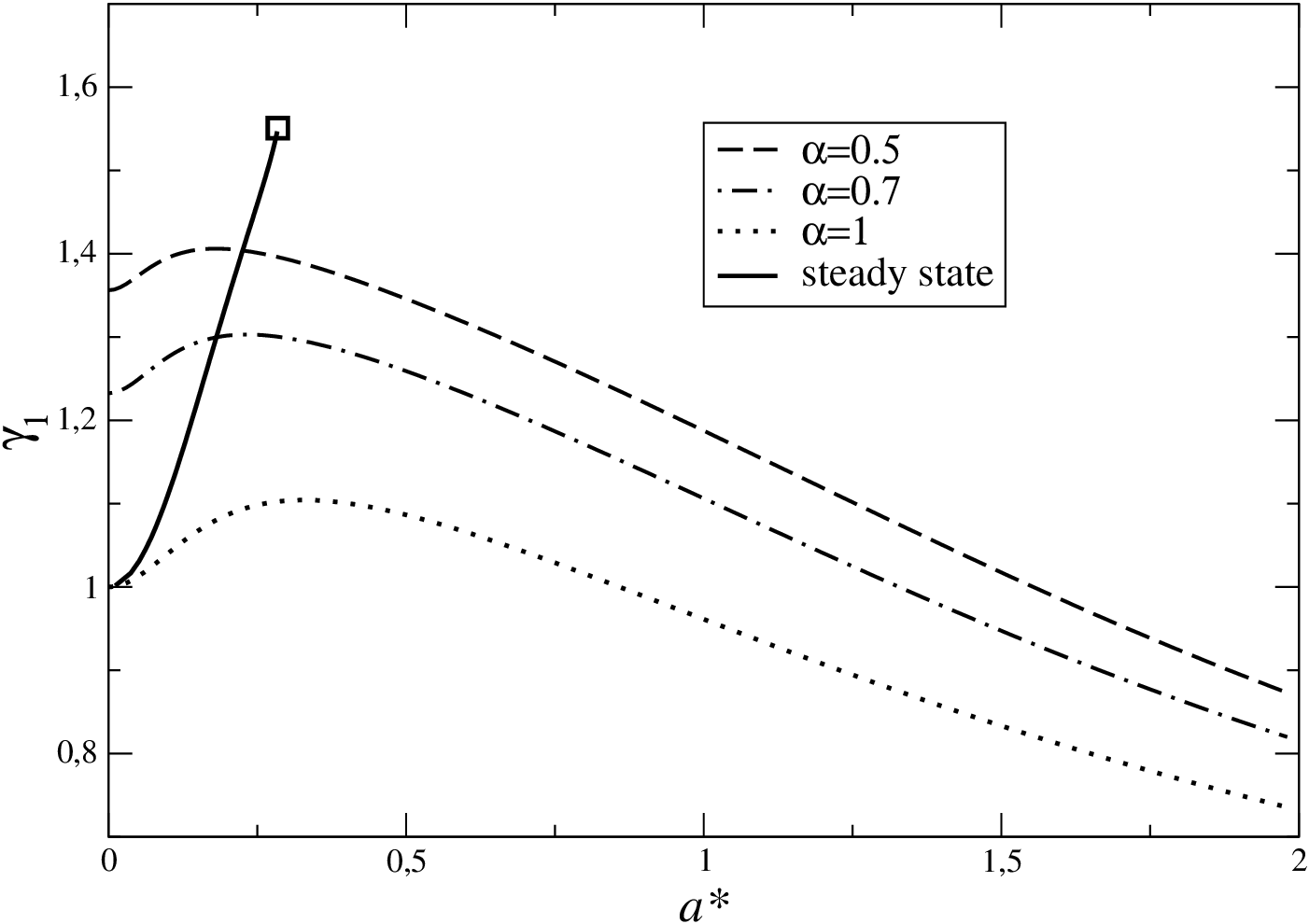}
\end{minipage}
\null\vskip 2mm\caption{Shear-rate dependence of the temperature ratio $\gamma_1=T_1/T$
in three dimensions ($d=3$), for an equimolar mixture ($x_1=0.5$), with $\sigma_1/\sigma_2=2$,
and $m_1/m_2=8$.  Three values of the (common) coefficient of restitution are displayed, together with the stationary curve, also shown in Figure \ref{fig:statioA} (see text for details).}
\label{fig:A1}
\null\vskip 5mm
\end{figure}

\begin{figure}[!h]
\null\vskip 6mm
\begin{tabular}{ccc}
\includegraphics[width=0.5\textwidth]{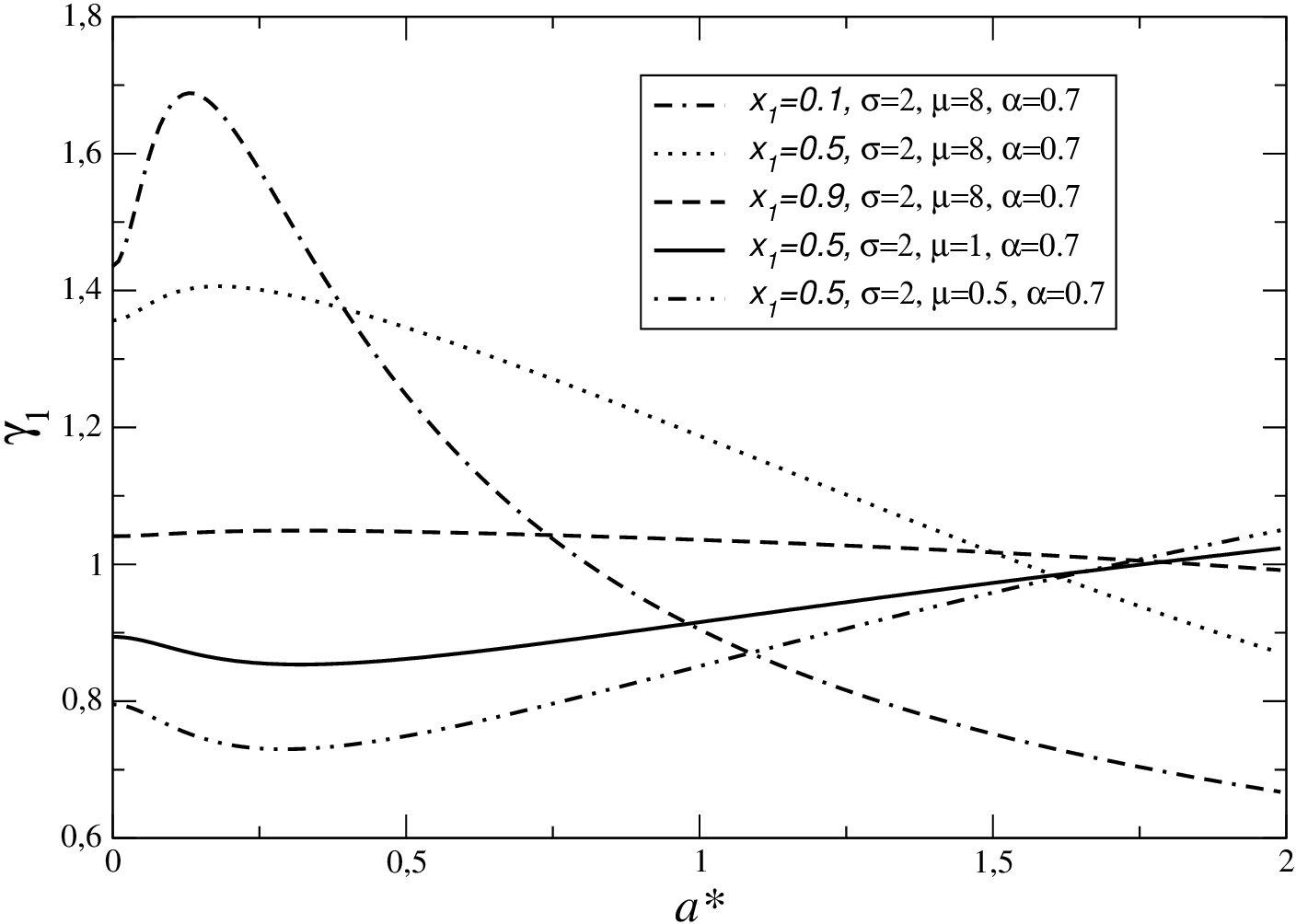}
\end{tabular}
\null\vskip 2mm\caption{Temperature ratio $\gamma_1=T_1/T$ as a function of the reduced shear rate
$a^*$ in three dimensions ($d=3$). Here, $\sigma$ denotes
the size ratio $\sigma_1/\sigma_2$ while $\mu\equiv m_1/m_2$.}
\label{fig:A3a}
\end{figure}

The expressions for the rheological functions $\eta^*(a^*)$ and $\Psi_1(a^*)$ depend on many parameters: $\left\{x_1, m_1/m_2, \sigma_1/\sigma_2, \alpha_{11}, \alpha_{22}, \alpha_{12}, a^*\right\}$. Obviously, this complexity exists in the elastic case as well \cite{MGS96},
so that the primary new feature is the dependence of $\eta^*(a^*)$ and $\Psi_1(a^*)$ on dissipation, on which
we shall concentrate. Also, for simplicity, we take the simplest case of common coefficient of restitution ($\alpha_{11}=\alpha_{22}=\alpha_{12}\equiv\alpha$). This reduces the parameter space to five quantities: $\left\{x_1, m_1/m_2, \sigma_1/\sigma_2, \alpha, a^*\right\}$. Before considering the rheological functions $\eta^*(a^*)$ and $\Psi_1(a^*)$, it is interesting to analyze the dependence of the temperature ratio $T_1/T_2$ on the shear rate. This quantity measures the lack of equipartition of the kinetic energy. Obviously, $T_1=T_2$ for any value of the shear rate and/or the coefficient of restitution in the case of mechanically equivalent particles. Figure \ref{fig:A1}
shows the temperature ratio as a function of the (reduced) shear rate $a^*$ in a situation where the grains have the same mass
density [$(\sigma_1/\sigma_2)^d=m_1/m_2$]. As is often the case in driven binary granular gases, the more massive particles have a larger granular temperature
for moderate shear rates, while the reverse conclusion holds at high shears.
The inelasticity parameter $\alpha$ and shear rate $a^*$ are here considered as independent, which in general results
in an unsteady situation for the system. The intersection of a curve $\gamma_1(a^*)$ for a given
$\alpha$ with the steady state line shown by the continuous thick line, provides the shear rate $a^*_s$ corresponding
to an exact balance between viscous heating and inelastic dissipation. For $a^* > a^*_s$, the temperature diverges,
while it decays to 0 in the opposite case. We note that even in the elastic case, the temperature
ratio differs from unity, as a signature of non equilibrium behaviour. Only when the shear rate does vanish
do we recover the equilibrium equipartition result ($\gamma_1=1$).
Fixing dissipation at $\alpha=0.7$, Fig. \ref{fig:A3a} complements the results
of Fig. \ref{fig:A1} by showing the
influence of mixture composition $x_1$. The same qualitative trends are observed as in Fig.
\ref{fig:A1}, see the three upper curves. However, Fig. \ref{fig:A3a} also shows that changing
the mass ratio (other parameters being fixed) can alter the results and lead to an increasing
ratio $T_1/T$ with increasing $a^*$.

\null\vskip 5mm
\begin{figure}[!h]
\begin{minipage}[c]{\textwidth}
\null\vskip 5mm
\includegraphics[width=0.5\textwidth]{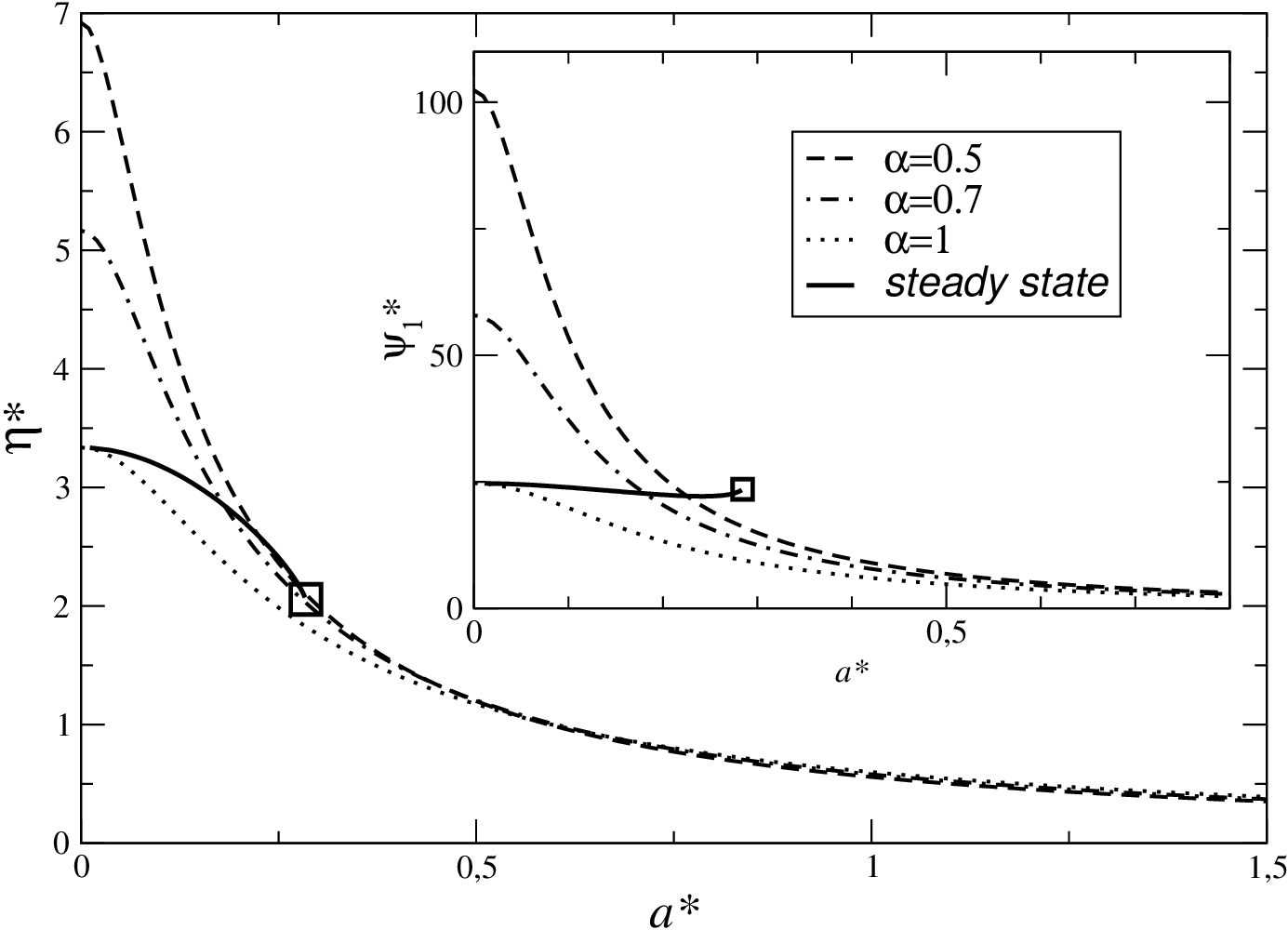}
\end{minipage}
\null\vskip 2mm\caption{Shear-rate dependence of the reduced non-linear
shear viscosity $\eta^*$
for $d=3$, an equimolar mixture ($x_1=0.5$), $\sigma_1/\sigma_2=2$,
and $m_1/m_2=8$, for three values of the (common) coefficient of restitution $\alpha$
(same situation as in Fig. \ref{fig:A1}).
The inset shows $\Psi_1^*$
versus the reduced shear rate $a^*$. The continuous thick curves
are the loci of steady-state results (also shown in Fig.
\ref{fig:statioA}) when dissipation is
scanned in the admissible range $\alpha \in [0,1]$. The squares
locate the terminal shear rate at maximum dissipation.}
\label{fig:A2}
\end{figure}

\begin{figure}[!h]
\null\vskip 6mm
\begin{tabular}{ccc}
\includegraphics[width=0.5\textwidth]{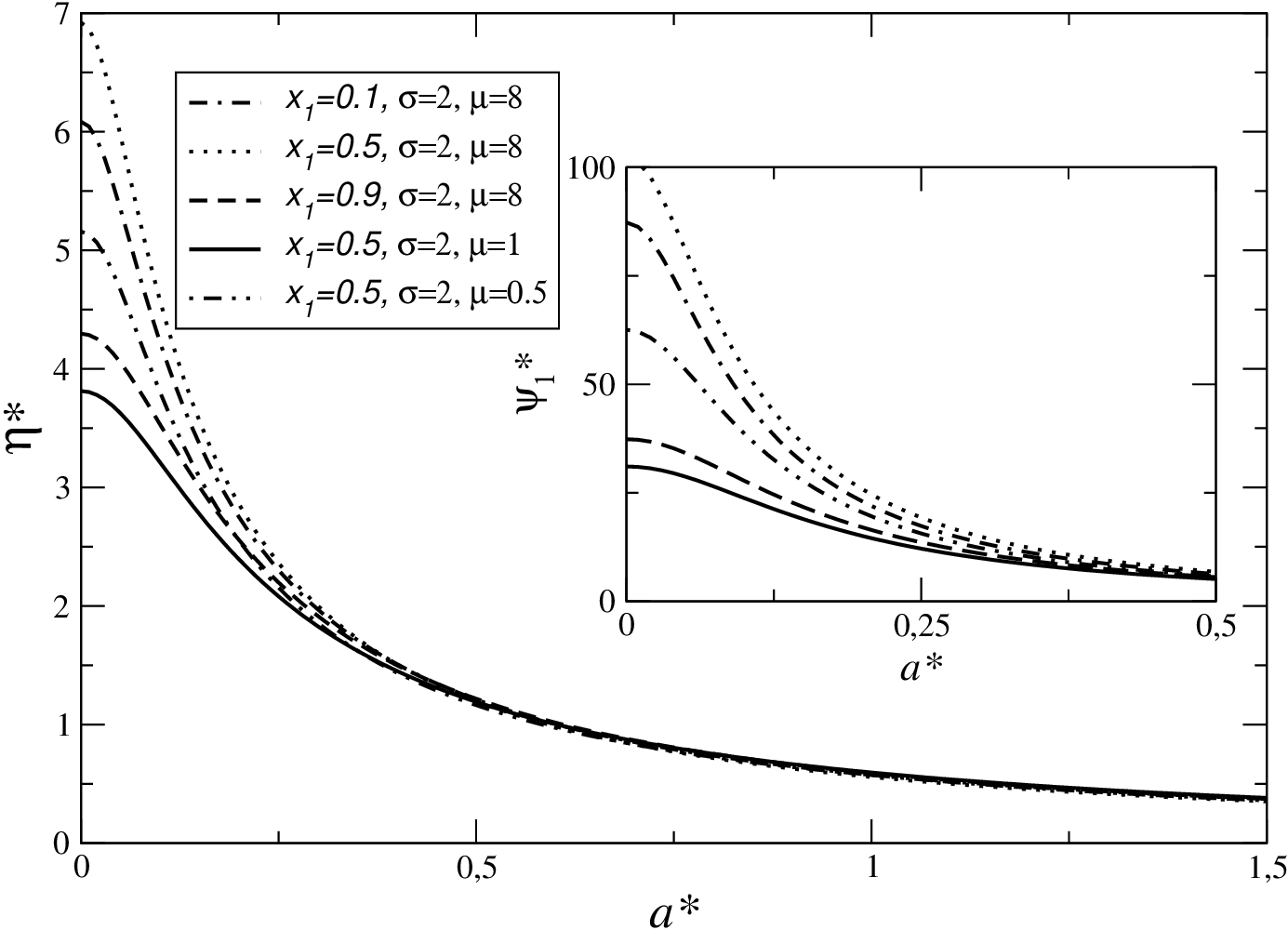}
\end{tabular}
\null\vskip 2mm\caption{Non-linear shear viscosity $\eta^*$ and normal stress difference $\Psi_1^*$ versus $a^*$.
The parameters are the same as for Fig. \ref{fig:A3a}, and the different curve
styles have the same meaning as in Fig. \ref{fig:A3a}.}
\label{fig:A3b}
\null\vskip 0mm
\end{figure}

\begin{figure}[ht]
\begin{minipage}[c]{\textwidth}
\null\vskip 5mm
\includegraphics[width=0.5\textwidth]{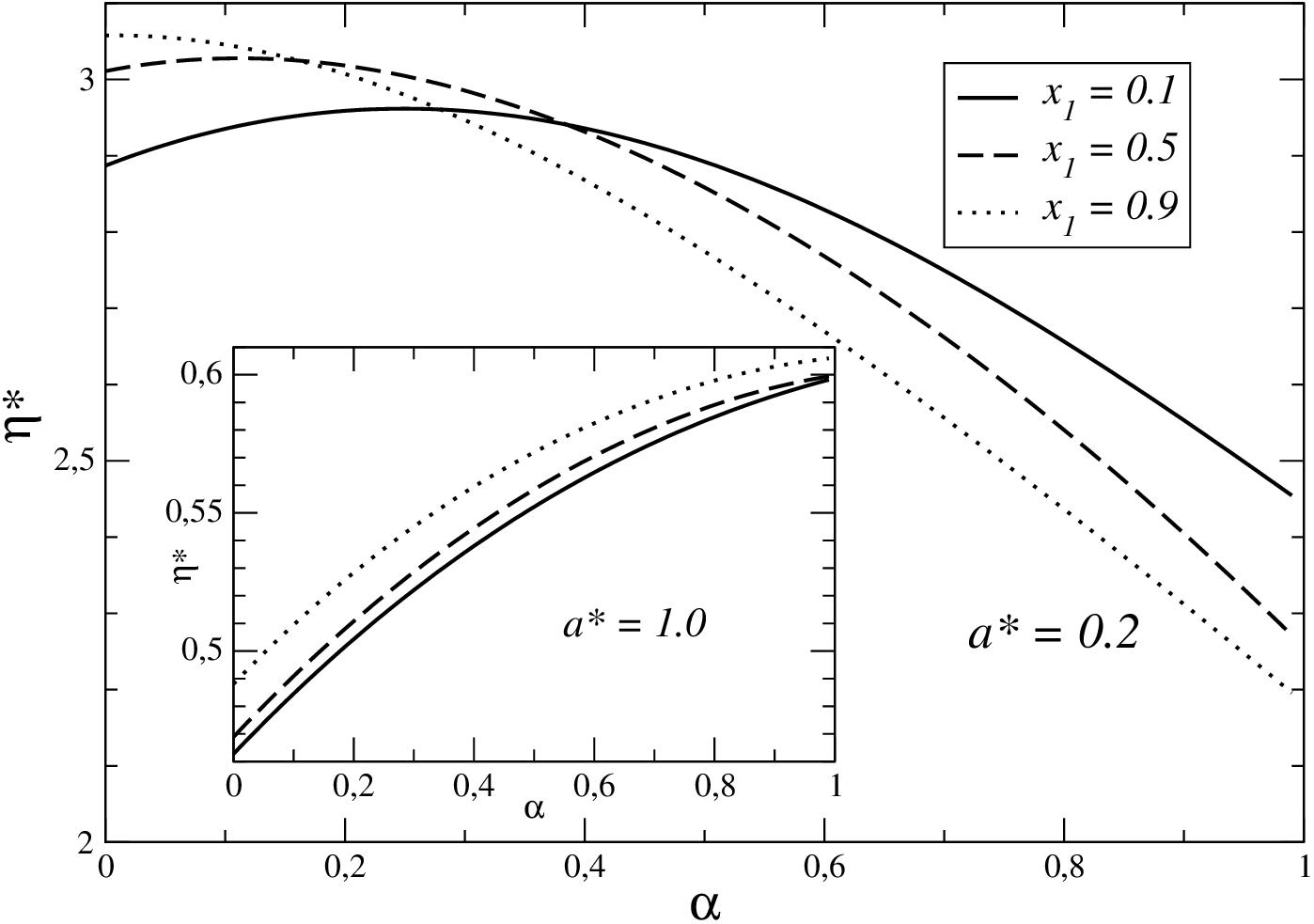}
\end{minipage}
\null\vskip 2mm\caption{Non-linear shear viscosity $\eta^*$
for $d=3$,  $\sigma_1/\sigma_2=2$,
and $\mu_1/\mu_2=8$, as a function of the (common) coefficient of restitution $\alpha$, for three mole fractions.
The main graph is for $a^*=0.2$ and the inset is for $a^*=1$.}
\label{fig:Aetavsalpha}
\end{figure}

\begin{figure}[!h]
\begin{minipage}[c]{\textwidth}
\null\vskip 5mm
\includegraphics[width=0.5\textwidth]{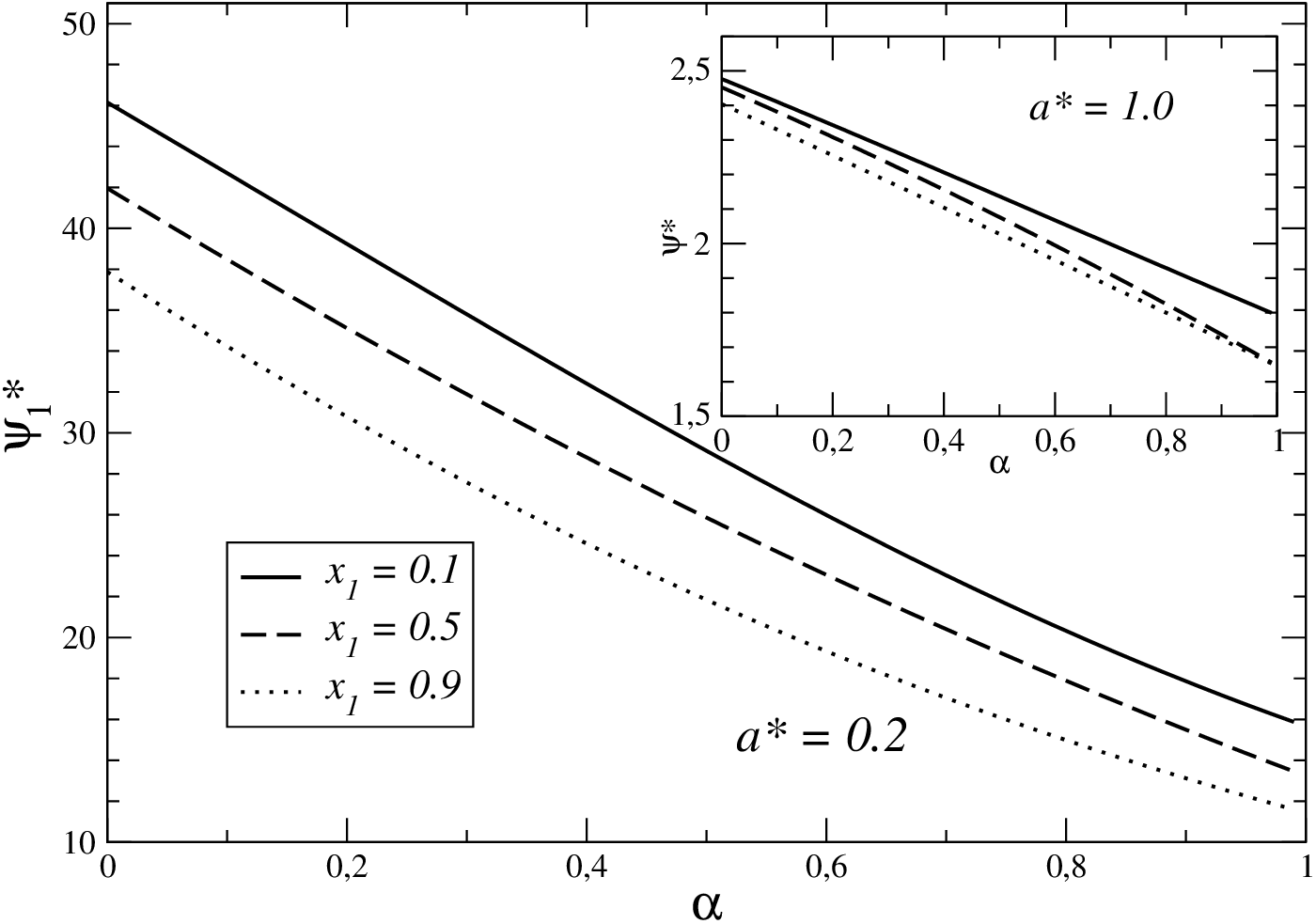}
\end{minipage}
\null\vskip 2mm\caption{Same as Fig. \ref{fig:Aetavsalpha}, but for the normal stress
difference $\Psi_1^*$.}
\label{fig:Apsivsalpha}
\end{figure}

It appears that the non-linear shear viscosity and viscometric function exhibit a more robust behaviour
with shear rate. It can be seen in Fig. \ref{fig:A2} that those functions decrease with increasing
$a^*$. In this figure, the steady state values are shown with the continuous thick curve, and the maximum
possible stationary shear rate is indicated by the squares (which correspond to $\alpha=0$).
As above, for a given inelasticity, the intersection of the $\eta^*$
(resp $\Psi_1^*$)
curve with its steady-state counterpart
determines the steady-state value of the shear rate
(resp normal stress difference).
As in Fig. \ref{fig:A1}, three coefficients of
restitution have been chosen in Fig. \ref{fig:A2}, and correspond to
strongly inelastic ($\alpha=0.5$), moderately inelastic
($\alpha=0.7$), and elastic systems ($\alpha=1$).
For completeness, we also show in Fig. \ref{fig:A3b} the rheological functions corresponding
to the parameter set of Fig. \ref{fig:A3a}. The same qualitative trend is observed as
in Fig. \ref{fig:A2}.
We note in Figures
\ref{fig:A3a}, \ref{fig:A2} and \ref{fig:A3b} the systematic trend that at high shear
rates, the shear viscosity and normal stress difference become
practically insensitive to the parameters specifying the
state of the system, in particular dissipation.
This observation is reminiscent of the single
species phenomenology \cite{SG07}.
We also observe that the dependence
on mixture composition $x_1$ and mass ratio is more subtle
(see e.g. the $\sigma_1/\sigma_2=2$ and $m_1/m_2=8$ curves showing that
the largest shear stress and normal stress difference occur in the equimolar case).
Likewise, it is observed that for $x_1=0.5$, $\sigma_1/\sigma_2=2$,
the smallest shear stress and normal stress difference
correspond to like masses ($m_1=m_2$).
We do not dwell on those effects since they are already present in the
vanishing shear rate limit. A non-trivial effect of shear rate, however,
is illustrated in Fig. \ref{fig:Aetavsalpha}: whereas at small $a^*$, the viscosity
function decreases with increasing $\alpha$ (at least in the physically relevant
range $\alpha>0.7$), increasing the shear rate leads to the opposite effect
(see the inset of Fig. \ref{fig:Aetavsalpha}). Enhanced dissipation then leads to smaller
shear stresses. In Fig. \ref{fig:Aetavsalpha}, we have displayed the full
possible range $0 \leq \alpha\leq 1$ to show that even at small shear rates,
an extreme dissipation can lead to a decreasing shear viscosity.
We finally note here that those effects are absent for the normal
stress difference, that appears to be a decreasing function
of $\alpha$, see Fig. \ref{fig:Apsivsalpha}.

\null\vskip 5mm

\subsection{Model B}

In Model B the collision frequency $\nu_0(T)$ is an increasing function of temperature, and so the
reduced shear rate $a^*$ is not constant. In order to have  $\eta^*(a^*)$ and $\Psi_1(a^*)$ in Model B,
one has to solve numerically the non-linear set (\ref{3.5}), discard the kinetic stage of the evolution,
and eliminate time in favor of $a^*(t)$ \cite{SGD04,SG07}. In addition, it should be remembered that
the above functions depend on the temperature ratio, that is itself time dependent
through its dependence on
$a^*(t)$. The above task
in the case of a mixture is therefore a significantly more complex problem than in the monodisperse case.
However, the results derived in the single gas case \cite{SG07} indicate that the influence of
the temperature dependence of $\nu_0$ on the rheological properties is quite small.
We then restrict here our discussion to the steady-state solution for Model B.
In this case, it is easy to see that the results obtained in the steady simple shear flow state are
\emph{universal} in the sense that they apply both for Model A and Model B, regardless of the specific
dependence of $\nu_0$ on $T$.

\begin{figure}[!h]
\begin{tabular}{ccc}
\includegraphics[width=0.45\textwidth]{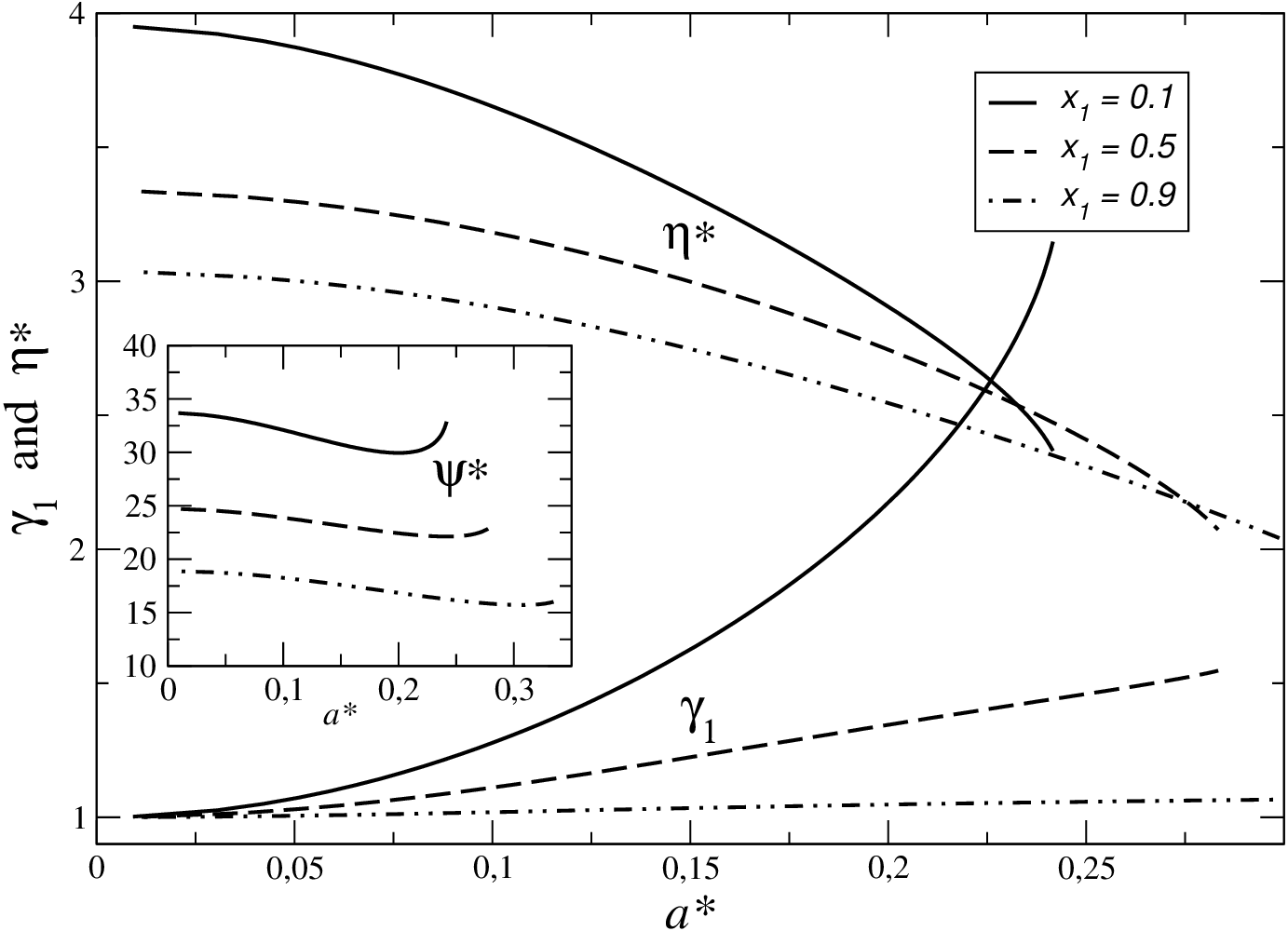} & ~~~~&
\includegraphics[width=0.455\textwidth]{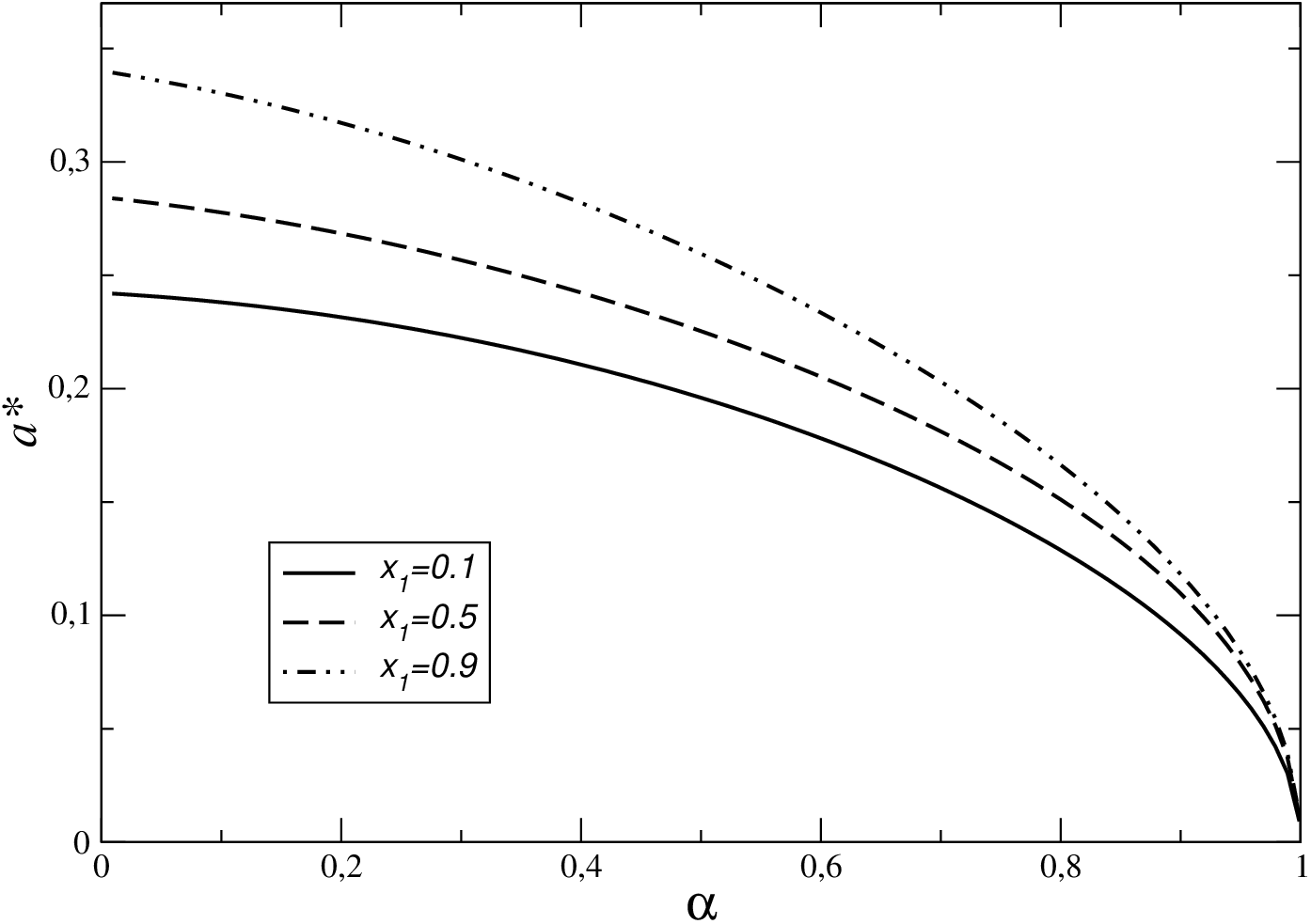}
\end{tabular}
\null\vskip 2mm\caption{(left) Plots of the temperature ratio $\gamma_1=T_1/T_2$ and the non-linear shear viscosity $\eta^*$ as functions of the (reduced)
shear rate $a^*$ in the steady state
for $d=3$, $\sigma_1/\sigma_2=2$, $m_1/m_2=8$, that corresponds to grains
of same mass density. Three different compositions are displayed.
The inset shows the normal stress difference $\psi^*\equiv \psi_1^*$ versus reduced shear rate. The right-hand side
graph shows the $\alpha$-dependence of the (reduced) shear rate $a^*$ for the above three systems.}
\label{fig:statioA}
\end{figure}

Figure \ref{fig:statioA} shows the corresponding rheological functions and temperature
ratio obtained at long times for a selected parameter set. The Figure also illustrates
that in the steady state, dissipation and reduced shear are coupled
(see the graph on the right hand side):
to every value of $a^*$ is associated a given inelasticity,
so that $a^*$ vanishes in the elastic limit $\alpha\to 1$. This explains why the
stationary temperature ratio converges to 1 in the small shear limit.
For a given set of parameters $m_1/m_2$, $\sigma_1/\sigma_2$, and mole
fraction $x_1$, the curves displayed are obtained by scanning all possible
inelasticity parameters $\alpha$ from 1 corresponding to a vanishing
stationary shear rate, to $\alpha=0$, which yields the maximum
possible value of $a^*$ (e.g. 0.24 for $x=0.1$ for the parameters
used in Fig. \ref{fig:statioA}, as can be seen in both left and right
hand side graphs).
The equimolar case results have already been shown as the continuous
curves in Figs. \ref{fig:A1} and \ref{fig:A2}. Interestingly,
it can be seen that the normal stress difference can become an increasing
function of $a^*$ (see the inset of Fig. \ref{fig:statioA}),
whereas it is decreasing within Model A, when $\alpha$ is fixed and
$a^*$ is changed. This is an illustration of the conflicting effects
at work at the stationary point, when simultaneously increasing $a^*$ and dissipation
(indeed, $\Psi_1^*$ decreases when $\alpha$ is fixed and $a^*$ increases while
it increases when $a^*$ is fixed and $\alpha$ decreases).
As far as the shear viscosity is concerned, the effects at work always lead to a decreasing
function of $a^*$, as with fixed dissipation within Model A.

\null\vskip 5mm

\section{Conclusions}
\label{sec5}

We have determined the rheological properties (shear stress and normal stress difference)
for a binary granular mixture in a uniform shear flow. The problem has been addressed
in the framework of the Boltzmann equation with Maxwell kernel. An important advantage of such a model
--compared to the more realistic inelastic hard sphere kernel--
is that the collisional moments can be obtained exactly, and do not require the explicit knowledge of
the velocity distribution function of both species (only low order moments are required).
It is important here to stress that in spite of its approximate nature, IMM
have been shown to fare very favorably against Monte Carlo simulations of inelastic
hard sphere mixtures \cite{G03,MG02}. We therefore expect the trends reported here
to be realistic and of practical interest.

In the case of IMM, there is a characteristic frequency $\nu_0$ that can be chosen
freely. To study the rheological properties, two classes of models have been introduced.
In Model A, $\nu_0$ is taken independent from the total kinetic temperature $T$,
while in Model B, $\nu_0$
scales like $T^{q}$, with $q=1/2$ to reproduce the hard-sphere behaviour.
In this respect, Model A fundamentally differs from models with $q\neq 0$ in that
it allows to disentangle the effects of dissipation (embodied in the
coefficients of restitution  $\alpha_{rs}$) from those of imposed shear (embodied in the
reduced shear rate $a^*$ defined from the actual shear rate $a$ by $a^*=a/\nu_0$).
Indeed, whenever $q\neq 0$, the system reaches at long times a steady state
where viscous heating compensates for collisional cooling. In this steady state,
a key point is that the reduced shear rate is enslaved to the inelasticity, so that
the problem is inherently non-Newtonian \cite{SGD04} (in other words, it is not possible
to decrease the reduced shear rate $a^*$ by decreasing $a$, since then collisional dissipation
will be more efficient and lead to a smaller granular temperature, so that $a^*$
is finally unaffected). On the other hand, within Model A, inelastic dissipation and
viscous heating {\em generically} do not compensate, so that the temperature of the system
either grows without bounds, or decreases to 0. If, however, the reduced shear rate is
adjusted to the precise value that is reached within Model B for a given parameter set,
Model A admits a steady state (which is then identical to its Model B counterpart).

Outside the particular steady state point and within Model A, the scaled pressure tensor
reaches well-defined stationary values which are non-linear functions of
$a^*$ and the coefficients of restitution $\alpha_{rs}$. This allows for a clear-cut definition
of the (reduced) non-linear shear viscosity $\eta^*$ and normal stress difference $\Psi_1^*$
(within Maxwell models, only one stress difference is non-vanishing, whereas there
are two such quantities for the hard sphere kernel). These quantities have
been computed for various parameters characterizing the mixture, and it has been
found that when $a^*$ is large enough (say $a^*>1$), $\eta^*$ and $\Psi_1^*$
are insensitive to dissipation, mixture composition, mass ratio, and size ratio, but only depend
on the shear rate. Both quantities decrease when $a^*$ increases. At smaller
shear amplitudes, the detailed parameters of the mixture becomes relevant
and in particular we have found --for physically relevant dissipation parameters--
that the shear viscosity and normal stress difference
increase when dissipation is increased. The dependence on mixture composition and mass ratio
appear more subtle, and are non-monotonous. Of particular interest also is
the temperature ratio, which has been seen to
depend on the parameters of the problem in a complex fashion.

Future developments include the study of the tracer limit, segregation of an intruder
by thermal diffusion, together with a
more general derivation of generalized transport coefficients from a Chapman-Enskog-like
expansion \cite{G06}. Work along these lines is in progress.

\acknowledgments

This work was initiated during a visit of V.G. to the Laboratoire de Physique
Thé\'eorique et Mod\`eles Statistiques, Universit\'e Paris-Sud. He is grateful to this institution
for its hospitality and support. The research of V.G. has been supported by the Ministerio de
Educaci\'on y Ciencia (Spain) through grant No. FIS2007-60977, partially financed by
FEDER funds and by the Junta de Extremadura (Spain) through Grant No. GRU08069.

\appendix
\section{Partial pressure tensors}

The expression of the relevant elements of the partial pressure tensor ${\sf P}_1^*$ are given by
\begin{widetext}
\begin{eqnarray}
\label{a1}
P_{1,xx}^*&=&\frac{1}{\Delta^3}\left\{G\Delta^2 B_{12}^*+2 a^{*2}G B_{12}\left[3\lambda^2+
B_{11}^{*2}+B_{12}^*B_{21}^*+B_{22}^*(B_{22}^*+3\lambda)+B_{11}^*(B_{22}^*+3\lambda)\right]\right.\nonumber\\
& & \left. -F \Delta^2(B_{22}^*+\lambda)-2 a^{*2} F \left[B_{11}^*B_{12}^*B_{21}^*
+B_{12}^*B_{21}^*(2B_{22}^*+3\lambda)+(B_{22}^*+\lambda)^3\right]\right\},
\end{eqnarray}
\end{widetext}
\begin{equation}
\label{a2} P_{1,yy}^*=\frac{G B_{12}-F(B_{22}+\lambda)}{\Delta},
\end{equation}
\begin{equation}
\label{a3} P_{1,xy}^*=-\frac{a^*}{\Delta^2}\left\{F\left[B_{12}B_{21}+(B_{22}+\lambda)^2\right]-G B_{12}\left(B_{11}+B_{22}+2\lambda\right)\right\},
\end{equation}
where
\begin{equation}
\label{a4}
\Delta=B_{12}^*B_{21}^*-(B_{11}^*+\lambda)(B_{22}^*+\lambda),
\end{equation}
\begin{equation}
\label{a5}
F=A_{11}^*p_1^*+A_{12}p_2^*,
\end{equation}
\begin{equation}
\label{a6} G=A_{22}^*p_2^*+A_{21}p_1^*.
\end{equation}
In the above equations, $A_{rs}^*=A_{rs}/\nu_0$ and $B_{rs}^*=B_{rs}/\nu_0$
where $A_{rs}$ and $B_{rs}$ are given by Eqs.\ (\ref{3.6})--(\ref{3.9}).
The expressions for $P_{2,ij}^*$ can easily obtained from (\ref{a1})--(\ref{a3}) by the adequate changes of indices.


\begin{thebibliography} {99}


\bibitem{CC70}S. Chapman, T. G. Cowling, The Mathematical Theory of Nonuniform Gases,
Cambridge University Press, Cambridge, 1970.

\bibitem{BDKS98}J. J. Brey, J. W. Dufty, C. S. Kim, A. Santos,
Hydrodynamics for granular flow at low-density, Phys. Rev. E 58 (1998) 4638--4653;
V. Garz\'o, J. M. Montanero, Transport coefficients of a heated granular gas,
Physica A 313 (2002) 336--356.

\bibitem{IHS}See for instance, C. K. K. Lun, S. B. Savage, D. J. Jeffrey, N. Chepurniy,
Kinetic theories for granular flow: inelastic particles in Couette flow and
slightly inelastic particles in a general flow field, J. Fluid Mech. 140 (1984) 223--256; J. T. Jenkins, M. W. Richman, Plane simple shear of smooth inelastic circular disks: the anisotropy of the second moment in the dilute and dense limits, J. Fluid Mech. 192 (1988) 313--328; N. Sela, I. Goldhirsch, S. H. Noskowicz, Kinetic theoretical study of a simple sheared two-dimensional granular gas to Burnett order, Phys. Fluids 8 (1996) 2337--2353; J. J. Brey, M. J. Ruiz-Montero, F. Moreno, Steady uniform shear flow in a low density granular gas,  Phys. Rev. E 55 (1997) 2846--2856; J. M. Montanero, V. Garz\'o, A. Santos, J. J. Brey, Kinetic theory of simple granular shear flows of smooth hard spheres, J. Fluid. Mech. 389 (1999) 391--411;  V. Garz\'o, Tracer diffusion in granular shear flows, Phys. Rev. E 66 (2002) 021308; J. F. Lutsko, Rheology of dense polydisperse granular fluids under shear, Phys. Rev. E 70 (2004) 061101.

\bibitem{Maxwell}See for instance, A. V. Bobylev, J. A. Carrillo, I. M. Gamba, On some properties of kinetic and hydrodynamic equations for inelastic interactions, J. Stat. Phys. 98 (2000) 743--773; J. A. Carrillo, C. Cercignani, I. M. Gamba, Steady states of a Boltzmann equation for driven granular media, Phys. Rev. E 61 (2000) 7700--7707; E. Ben-Naim, P. L. Krapivsky, Multiscaling in inelastic collisions,  Phys. Rev. E 61 (2000) R5--R8; C. Cercignani, Shear flow of a granular material, J. Stat. Phys. 102 (2001) 1407--1415; M. H. Ernst, R. Brito, Scaling solutions of inelastic Boltzmann equations with over-populated high-energy tails, J. Stat. Phys. 109 (2002) 407--432; E. Ben-Naim, P. L. Krapivsky, The inelastic Maxwell model, Granular Gas Dynamics, in Lecture Notes in Physics 624, (2003) 65--94.

\bibitem{G03}V. Garz\'o, Nonlinear transport in inelastic Maxwell mixtures under simple shear flow, J. Stat. Phys. 112 (2003) 657--683.

\bibitem{TK03}E. Trizac, P. Krapivsky, Correlations in ballistic processes,
Phys. Rev. Lett. 91 (2003) 218302;  F. Coppex,   M. Droz, E. Trizac,
Maxwell and very hard particle models for probabilistic ballistic annihilation: Hydrodynamic description,
Phys. Rev. E 72 (2005) 021105.

\bibitem{SG07}A. Santos, V. Garz\'o, Simple shear flow in inelastic Maxwell models, J. Stat. Mech. P08021 (2007).



\bibitem{exp05}K. Kohlstedt, A. Snezhko, M. V. Sapozhnikov, I. S. Arnarson,
J. S. Olafsen, E. Ben-Naim, Velocity Distributions of granular gases with drag and with long-range interactions, Phys. Rev. Lett. 95 (2005) 068001.


\bibitem{Go03}I. Goldhirsch, Rapid granular flows, Annu. Rev. Fluid Mech. 35 (2003) 267-293.

\bibitem{SGD04}A. Santos, V. Garz\'o, J. W. Dufty, Inherent
rheology of a granular fluid in uniform shear flow, Phys.
Rev. E 69 (2004) 061303.

\bibitem{TM80}C. Truesdell, R. G. Muncaster, Fundamentals of Maxwell's Kinetic Theory of a Simple
Monatomic Gas, Academic Press, New York, 1980.

\bibitem{GS03}V. Garz\'o, A. Santos, Kinetic Theory of Gases in Shear Flows. Nonlinear Transport,
Kluwer Academic, Dordrecht, 2003.

\bibitem{ETB06}M.H. Ernst, E. Trizac, A. Barrat,  The rich behaviour of the Boltzmann equation for dissipative
gases, Europhys. Lett. 76 (2006) 56--62;
M. H. Ernst, E. Trizac, A. Barrat, The Boltzmann equation for driven systems of inelastic soft spheres, J. Stat. Phys. 124 (2006) 549--586; A. Barrat, E. Trizac, M. H. Ernst, Quasi-elastic solutions to the nonlinear Boltzmann equation for dissipative gases, J.Phys. A: Math. Theor. 40 (2007) 4057--4076.

\bibitem{computer}See for instance, J. M. Montanero, V. Garz\'o,
Monte Carlo simulation of the homogeneous cooling state for a granular mixture,
Granular Matter 4 (2002) 17--24; A.Barrat, E.Trizac, Molecular dynamics simulations of vibrated granular
gases, Phys. Rev. E 66 (2002) 051303;
A. Barrat, E. Trizac, Lack of energy equipartition in homogeneous heated binary granular mixtures,
Granular Matter 4 (2002) 57--63 (2002); S. R. Dahl, C. M. Hrenya, V. Garz\'o, J. W. Dufty, Kinetic temperatures
for a granular mixture, Phys. Rev. E 66 (2002) 041301; R. Pagnani, U. M. B. Marconi, and A. Puglisi,
Driven low density granular mixtures, Phys. Rev. E 66 (2002) 051304;  P. Krouskop, J. Talbot, Mass and size effects in three-dimensional vibrofluidized granular
mixtures, Phys. Rev. E 68 (2003) 021304; H. Wang, G. Jin, Y. Ma, Simulation study on kinetic
temperatures of vibrated binary granular mixtures, Phys. Rev. E 68 (2003) 031301; J. J. Brey, M. J.
Ruiz-Montero, F. Moreno, Energy partition and segregation for an intruder in a vibrated granular system
under gravity, Phys. Rev. Lett. 95 (2005) 098001; M. Schr\"oter, S. Ulrich , J. Kreft , J. B. Swift, H. L. Swinney, Mechanisms in the size segregation of a binary granular mixture, Phys. Rev. E 74 (2006) 011307.


\bibitem{exp}R. D. Wildman, D. J. Parker, Coexistence of two granular
temperatures in binary vibrofluidized beds, Phys. Rev. Lett. 88 (2002) 064301; K. Feitosa, N.
Menon, Breakdown of energy equipartition in a 2D binary vibrated granular gas, Phys. Rev. Lett. 88 (2002) 198301.

\bibitem{GD99}  V. Garz\'{o}, J. W. Dufty, Homogeneous cooling
state for a granular mixture, Phys. Rev. E 60 (1999) 5706--5713.

\bibitem{LE72}A. W. Lees, S. F. Edwards, The computer study of transport processes under extreme conditions, J. Phys. C 5 (1972) 1921--1929.

\bibitem{Tij01}M. Tij, E. E. Tahiri, J. M. Montanero, V. Garz\'o, A. Santos, Nonlinear Couette flow in a low density granular gas, J. Stat. Phys. 103 (2001) 1035--1068.

\bibitem{VGS08}F. Vega Reyes, V. Garz\'o, A. Santos, Impurity in a granular gas under nonlinear Couette flow, J. Stat. Mech. P09003 (2008).



\bibitem{GS07}V. Garz\'o, A. Santos, Third and fourth degree collisional moments for inelastic Maxwell models, J. Phys. A: Math. Theor. 40 (2007) 14927--14943.

\bibitem{GA05}
V. Garz\'o, A. Astillero, Transport coefficients for inelastic Maxwell mixtures, J. Stat. Phys. 118 (2005) 935--971.


\bibitem{MG02}J. M. Montanero, V. Garz\'o, Rheological properties of in a low-density granular mixture, Physica A 310 (2002) 17--38.

\bibitem{MGS96}C. Mar\'{\i}n, V. Garz\'o, A. Santos, Transport properties in a binary mixture under shear flow, Phys. Rev. E 52 (1995) 3812--3820.

\bibitem{rque}
For mechanically equivalent particles, an explicit comparison with the
results reported in \cite{SG07} requires a mapping of time scales
by a factor $(d+2)/2$.
Indeed, the reduced shear rate $a^*$ used here is related to the
same quantity appearing in \cite{SG07}, which we denote
$a^*_{\text{SG}}$, through $a^* = 2a^*_{\text{SG}}/(d+2)$.
The rheological functions are likewise affected: denoting again
with subscript SG the reduced shear viscosity and viscometric function
derived in \cite{SG07}, we have
$$
\eta^* = \frac{d+2}{2}\,\eta^*_{\text{SG}} \qquad \hbox{and}\qquad
\Psi_1^* =  \left(\frac{d+2}{2}\right)^2\,\Psi^*_{1,\text{SG}}.
$$


\bibitem{G06}V. Garz\'o, Transport coefficients for an inelastic gas around uniform shear flow: Linear
stability analysis, Phys. Rev. E 73 (2006) 021304; V. Garz\'o, Shear-rate dependent transport coefficients
for inelastic Maxwell models, J. Phys. A: Math. Theor. 40 (2007) 10729--10757.


\end{thebibliography}
\end{document}